\input harvmac
\input epsf

\mathchardef\varGamma="0100
\mathchardef\varDelta="0101
\mathchardef\varTheta="0102
\mathchardef\varLambda="0103
\mathchardef\varXi="0104
\mathchardef\varPi="0105
\mathchardef\varSigma="0106
\mathchardef\varUpsilon="0107
\mathchardef\varPhi="0108
\mathchardef\varPsi="0109
\mathchardef\varOmega="010A

\font\mbm = msbm10
\font\Scr=rsfs10
\def\bb#1{\hbox{\mbm #1}}

\def\scr#1{\hbox{\Scr #1}}

\def\Mt{{\kern1em\hbox{$\tilde{\kern-1em{\scr M}}$}}}
\def\At{{\kern1em\hbox{$\tilde{\kern-1em{\scr A}}$}}}
\def\Kt{{\kern1em\hbox{$\tilde{\kern-1em{\scr K}}$}}}

\font\sScr=rsfs7
\def\sscr#1{\hbox{\sScr #1}}

\Title{\vbox{\rightline{\tt hep-th/0307254} \rightline{CERN-TH/2003-165}}}
{\vbox{\centerline{Suppressing the Cosmological Constant}
\vskip .1in
\centerline{in Non-Supersymmetric Type I Strings}
}}

\centerline{Carlo Angelantonj and Ignatios
Antoniadis~\footnote{$^{\dagger}$}{On leave from {\it CPHT
{\'E}cole Polytechnique (UMR du CNRS 7644), F-91128 Palaiseau}}}
\medskip
\centerline{\it CERN -- Theory Division, CH-1211 Geneva 23}

\vskip 0.3in

\centerline{{\bf Abstract}}

\noindent
We construct non-supersymmetric type I string models which
correspond to consistent flat-space solutions of all classical equations
of motion. Moreover, the one-loop vacuum energy is naturally fixed by the
size of compact extra dimensions which, in the two-dimensional case,
can be lowered to a fraction of a millimetre.
This class of models has interesting non-abelian gauge groups and can
accommodate chiral fermions. In the large radius limit,
supersymmetry is recovered in the bulk, while D-brane excitations,
although non-supersymmetric,
exhibit Fermi-Bose degeneracy at all mass levels. We also give some
evidence for a suppression of higher-loop corrections to the
vacuum energy.

%\noindent
%PACS: 11.25.-w, 11.25.Mj

%\noindent
%Keywords: Open Strings.

\Date{July 2003} %replace this line by \draft  for preliminary versions
%or specify \draftmode at some point
%\draft%

%%%%%%%%%%%%%%%%%%%%%%%%%%%%%%%%%%%%%%%%%%%%%%%%%%%%%%%%%%%%%%%%%%%%%%

\lref\KachruHD{
S.~Kachru, J.~Kumar and E.~Silverstein,
``Vacuum energy cancellation in a non-supersymmetric string,''
Phys.\ Rev.\ D59 (1999) 106004[arXiv:hep-th/9807076].
%%CITATION = HEP-TH 9807076;%%
}

\lref\KachruPG{
S.~Kachru and E.~Silverstein,
``On vanishing two loop cosmological constants in non-supersymmetric
strings,'' JHEP 9901 (1999) 004
[arXiv:hep-th/9810129];
%%CITATION = HEP-TH 9810129;%%
R.~Iengo and C.~J.~Zhu,
``Evidence for non-vanishing cosmological constant in non-SUSY superstring
models,''
JHEP 0004 (2000) 028
[arXiv:hep-th/9912074].
%%CITATION = HEP-TH 9912074;%%
}

\lref\AngelantonjGM{
C.~Angelantonj, I.~Antoniadis and K.~Forger,
``Non-supersymmetric type I strings with zero vacuum energy,''
Nucl.\ Phys.\ B555 (1999) 116
[arXiv:hep-th/9904092].
%%CITATION = HEP-TH 9904092;%%
}

\lref\BlumenhagenUF{
R.~Blumenhagen and L.~G\"orlich,
``Orientifolds of non-supersymmetric, asymmetric orbifolds,''
Nucl.\ Phys.\ B551 (1999) 601
[arXiv:hep-th/9812158].
%%CITATION = HEP-TH 9812158;%%
}

\lref\HarveyRC{
J.~A.~Harvey,
``String duality and non-supersymmetric strings,''
Phys.\ Rev.\ D59 (1999) 026002
[arXiv:hep-th/9807213].
%%CITATION = HEP-TH 9807213;%%
}

\lref\BurgessQW{
C.~P.~Burgess, R.~C.~Myers and F.~Quevedo,
``A naturally small cosmological constant on the brane?,''
Phys.\ Lett.\ B495 (2000) 384
[arXiv:hep-th/9911164];
%%CITATION = HEP-TH 9911164;%%
Y.~Aghababaie, C.~P.~Burgess, S.~L.~Parameswaran and F.~Quevedo,
``Toward a naturally small cosmological constant from branes in 6D
supergravity,'' [arXiv:hep-th/0304256].
%%CITATION = HEP-TH 0304256;%%
}
\lref\ItoyamaRC{
H.~Itoyama and T.~R.~Taylor,
``Supersymmetry Restoration In The Compactified O(16) X O(16)-Prime
Heterotic String Theory,'' Phys.\ Lett.\ B186 (1987) 129.
%%CITATION = PHLTA,B186,129;%%
}

\lref\AntoniadisZG{
I.~Antoniadis, S.~Dimopoulos and G.~R.~Dvali,
``Millimetre range forces in superstring theories with weak-scale
compactification,''Nucl.\ Phys.\ B516 (1998) 70
[arXiv:hep-ph/9710204].
%%CITATION = HEP-PH 9710204;%%
}

\lref\PolchinskiMT{
J.~Polchinski,
``Dirichlet-Branes and Ramond-Ramond Charges,''
Phys.\ Rev.\ Lett.\  75 (1995) 4724
[arXiv:hep-th/9510017].
%%CITATION = HEP-TH 9510017;%%
}
\lref\AntoniadisGW{
I.~Antoniadis, K.~Benakli, A.~Laugier and T.~Maillard,
``Brane to bulk supersymmetry breaking and radion force at micron
distances,'' Nucl.\ Phys.\ B662 (2003) 40
[arXiv:hep-ph/0211409].
%%CITATION = HEP-PH 0211409;%%
}
\lref\AngelantonjJH{
C.~Angelantonj,
``Comments on open-string orbifolds with a non-vanishing $B_{ab}$,''
Nucl.\ Phys.\ B566 (2000) 126
[arXiv:hep-th/9908064].
%%CITATION = HEP-TH 9908064;%%
}
\lref\BianchiEU{
M.~Bianchi, G.~Pradisi and A.~Sagnotti,
``Toroidal compactification and symmetry breaking in open string theories,''
Nucl.\ Phys.\ B376 (1992) 365.
%%CITATION = NUPHA,B376,365;%%
}

\lref\AngelantonjCT{
C.~Angelantonj and A.~Sagnotti,
``Open strings,''
Phys.\ Rept.\ 371 (2002) 1
[Erratum-ibid.\ 376 (2003) 339]
[arXiv:hep-th/0204089].
%%CITATION = HEP-TH 0204089;%%
}

\lref\AntoniadisKI{
I.~Antoniadis, E.~Dudas and A.~Sagnotti,
``Supersymmetry breaking, open strings and M-theory,''
Nucl.\ Phys.\ B544 (1999) 469
[arXiv:hep-th/9807011];
%%CITATION = HEP-TH 9807011;%%
I.~Antoniadis, G.~D'Appollonio, E.~Dudas and A.~Sagnotti,
``Partial breaking of supersymmetry, open strings and M-theory,''
Nucl.\ Phys.\ B553 (1999) 133
[arXiv:hep-th/9812118].
%%CITATION = HEP-TH 9812118;%%
}

\lref\SagnottiTW{
A.~Sagnotti,
``Open Strings And Their Symmetry Groups,''
Cargese Summer Institute on Non-Perturbative Methods in Field Theory,
Cargese, France, 1987
[arXiv:hep-th/0208020];
%%CITATION = HEP-TH 0208020;%%
M.~Bianchi and A.~Sagnotti,
``On The Systematics Of Open String Theories,''
Phys.\ Lett.\ B247 (1990) 517
%%CITATION = PHLTA,B247,517;%%
and
``Twist Symmetry And Open String Wilson Lines,''
Nucl.\ Phys.\ B361 (1991) 519;
%%CITATION = NUPHA,B361,519;%%
P.~Horava,
``Strings On World Sheet Orbifolds,''
Nucl.\ Phys.\ B327 (1989) 461.
%%CITATION = NUPHA,B327,461;%%
}

\lref\DudasBN{
E.~Dudas,
``Theory and phenomenology of type I strings and M-theory,''
Class.\ Quant.\ Grav.\ 17 (2000) R41
[arXiv:hep-ph/0006190].
%%CITATION = HEP-PH 0006190;%%
}

\lref\AntoniadisAX{
I.~Antoniadis and C.~Bachas,
``Branes and the gauge hierarchy,''
Phys.\ Lett.\ B450 (1999) 83
[arXiv:hep-th/9812093];
%%CITATION = HEP-TH 9812093;%%
N.~Arkani-Hamed, S.~Dimopoulos and J.~March-Russell,
``Stabilisation of sub-millimetre dimensions: The new guise of the
hierarchy problem,''
Phys.\ Rev.\ D63 (2001) 064020
[arXiv:hep-th/9809124].
%%CITATION = HEP-TH 9809124;%%
}

\lref\marco{
M.~Borunda, M.~Serone and M.~Trapletti,
``On the quantum stability of type IIB orbifolds and orientifolds 
with Scherk-Schwarz SUSY breaking,''
Nucl.\ Phys.\ B 653 (2003) 85
[arXiv:hep-th/0210075].
%%CITATION = HEP-TH 0210075;%%
}

\lref\SugimotoTX{
S.~Sugimoto,
``Anomaly cancellations in type I D9-D9-bar system and the USp(32)
string theory,''
Prog.\ Theor.\ Phys.\ 102 (1999) 685
[arXiv:hep-th/9905159];
%%CITATION = HEP-TH 9905159;%%
I.~Antoniadis, E.~Dudas and A.~Sagnotti,
``Brane supersymmetry breaking,''
Phys.\ Lett.\ B464 (1999) 38
[arXiv:hep-th/9908023];
%%CITATION = HEP-TH 9908023;%%
G.~Aldazabal and A.~M.~Uranga,
``Tachyon-free non-supersymmetric type IIB orientifolds via
brane-antibrane systems,''
JHEP 9910 (1999) 024
[arXiv:hep-th/9908072].
%%CITATION = HEP-TH 9908072;%%
}

\lref\WittenBS{
E.~Witten,
``Toroidal compactification without vector structure,''
JHEP 9802 (1998) 006
[arXiv:hep-th/9712028].
%%CITATION = HEP-TH 9712028;%%
}

\lref\SSi{
J.~Scherk and J.~H.~Schwarz,
``How To Get Masses From Extra Dimensions,''
Nucl.\ Phys.\ B153 (1979) 61
%%CITATION = NUPHA,B153,61;%%
and
``Spontaneous Breaking Of Supersymmetry Through Dimensional Reduction,''
Phys.\ Lett.\ B82 (1979) 60.
%%CITATION = PHLTA,B82,60;%%
}

\lref\SSiii{
R.~Rohm,
``Spontaneous Supersymmetry Breaking In Supersymmetric String Theories,''
Nucl.\ Phys.\ B237 (1984) 553;
%%CITATION = NUPHA,B237,553;%%
C.~Kounnas and M.~Porrati,
``Spontaneous Supersymmetry Breaking In String Theory,''
Nucl.\ Phys.\ B310 (1988) 355;
%%CITATION = NUPHA,B310,355;%%
S.~Ferrara, C.~Kounnas, M.~Porrati and F.~Zwirner,
``Superstrings With Spontaneously Broken Supersymmetry And Their
Effective Theories,''
Nucl.\ Phys.\ B318 (1989) 75.
%%CITATION = NUPHA,B318,75;%%
}

\lref\Ldim{
I.~Antoniadis, ``A Possible New Dimension at a Few TeV,''
Phys. Lett. B246 (1990)~377.
%%CITATION = PHLTA,B246,377;%%
}

\lref\ADD{
N.~Arkani-Hamed, S.~Dimopoulos and G.~Dvali,
``The Hierarchy Problem and New Dimensions at a Millimetre,''
Phys. Lett. B429 (1998) 263, [arXiv:hep-ph/9803315] and
``Phenomenology, astrophysics and cosmology of theories with
sub-millimetre dimensions and TeV scale quantum gravity,'' Phys.\ Rev.\
D59 (1999) 086004 [arXiv:hep-ph/9807344];
%%CITATION = HEP-PH 9807344;%%
I.~Antoniadis, N.~Arkani-Hamed, S.~Dimopoulos and G.~R.~Dvali,
``New dimensions at a millimetre to a Fermi and superstrings at a TeV,''
Phys.\ Lett.\ B436 (1998) 257
[arXiv:hep-ph/9804398].
%%CITATION = HEP-PH 9804398;%%
}

\lref\sp{
E.~Silverstein, private communication.
}

\lref\Cotrone{
A.~L.~Cotrone,
``A $\bb{Z}_2 \times \bb{Z}_2$ orientifold with spontaneously broken 
supersymmetry,''
Mod.\ Phys.\ Lett.\ A 14 (1999) 2487
[arXiv:hep-th/9909116].
%%CITATION = HEP-TH 9909116;%%
}

%%%%%%%%%%%%%%%%%%%%%%%%%%%%%%%%%%%%%%%%%%%%%%%%%%%%%%%%%%%%%%%%%%%%%%%%%

\newsec{Introduction}

One of the outstanding problems in string theory, as in any quantum
theory including gravity, is to understand the smallness of the
cosmological constant. This problem became much more severe after the
recent observations suggesting a non-vanishing value corresponding to a
new energy scale, by far smaller than every other scale in physics of
fundamental interactions, $\varLambda\equiv E_\varLambda^4$ with
$E_\varLambda\sim 10^{-4}\, {\rm eV}$. On the other hand,
brane-world models with large extra dimensions have recently received
a lot of attention, as they offer a new framework for stabilising the mass
hierarchy and addressing several problems of particle physics in an
interesting new perspective \refs{\Ldim, \ADD}.
In particular,  models with two
large flat dimensions of submillimeter size are particularly attractive
for manifold reasons: technical (maximum transverse space
available in phenomenologically interesting string constructions);
phenomenological (enhanced signals both in particle colliders and in
micro-gravity experiments); and theoretical, related mainly to the
logarithmic growth of massless field propagation in the bulk
\refs{\AntoniadisAX, \AntoniadisGW}, offering new
ways for addressing gauge coupling unification, neutrino masses and
oscillations, fixing the radion, etc.

It is then suggestive, if not a simple numerical coincidence, that the
size of the bulk $R_\perp$ in these models is of the same
order of magnitude as the
cosmological constant scale $R_\perp\sim E_\varLambda^{-1}\sim 1\,
{\rm mm}$\footnote{$^{\star}$}{See \BurgessQW\ where similar
ideas for suppressing the cosmological constant have been exposed, though
employing different methods.}.  In fact,
the behaviour $\varLambda\sim 1/R_\perp^4$ is generically valid in models
where supersymmetry is broken by Scherk-Schwarz boundary conditions
\refs{\SSi, \SSiii, \AntoniadisKI, \Cotrone, \ItoyamaRC, \Ldim,
\AntoniadisZG}. However, in this case the induced mass splittings are of
order $1/R_\perp$, and thus too small by several orders of magnitude.
It would then be very appealing if supersymmetry might be broken
on the branes at the string
scale $M_s\sim 1\, {\rm TeV}$, in a way that the vacuum energy vanishes in the
decompactification limit. Note that in this limit four-dimensional
gravity decouples, since the effective Planck mass $M_P\simeq M_s^2
R_\perp$ goes to infinity.

Type II string models with possible vanishing cosmological constant in
perturbation theory were studied in refs. \refs{\KachruHD, \HarveyRC}.
Their main feature is a Fermi-Bose degenerate spectrum, thus
leading to an automatic vanishing of the one-loop vacuum energy.
Aside from the question of possible higher-loop corrections \refs{\KachruPG},
their main defect, however, was that the non-abelian gauge
sector, appearing at particular singular points of the compactification
manifold, or on appropriate D-brane collections, was always
supersymmetric \refs{\BlumenhagenUF, \AngelantonjGM}.
Thus, it is questionable whether such constructions can
accommodate gauge degrees of freedom with large supersymmetry-breaking
mass splittings.

In this work, we address this problem and construct type I string
models with non-supersymmetric D-brane spectra and an additional source
of supersymmetry breaking in the bulk induced by Scherk-Schwarz boundary
conditions. In the large radius limit, the one-loop cosmological constant
vanishes as $1/R_\perp^4$, while the D-brane spectra exhibit
fermion-boson degeneracy at all massive levels. In the simplest case, this
degeneracy can be thought of as emerging from a supersymmetric theory, where
superpartners have been displaced appropriately in the position space.

Our paper is organised as follows. In section 2, we present a brief
review of the so-called M-theory breaking mechanism,
exhibiting either ``brane
supersymmetry'' or ``brane supersymmetry breaking''. In section 3, we
describe how discrete deformations can trade ${\scr O}^+$ planes for
${\scr O}^-$ ones. In section 4,
we construct the a simple model exhibiting open-string Fermi-Bose
degeneracy in the large-radius limit, at all mass levels.
The one-loop vacuum energy is computed in section 5,
where we show the previously announced behaviour.
In section 6, we generalise the construction to  lower dimensions.
Finally, in section 7, we comment on the possibility that higher-order
corrections do not destabilise the one-loop result.

\newsec{M-theory breaking revisited}

In order to make our analysis more explicit,
it is worth to review in this section the basic features of the ``M-theory
breaking'' mechanism, exhibiting the phenomena of either ``brane
supersymmetry'' \AntoniadisKI\ or ``brane supersymmetry breaking''
\refs{\SugimotoTX, \AntoniadisGW}.

In oriented closed-string theories, the Scherk-Schwarz reduction results
from a discrete deformation compatible with modular invariance
\refs{\SSiii}.
The general method for breaking supersymmetry by compactification uses a
(discrete) R-symmetry of the higher-dimensional theory and couples the lattice
momenta to the corresponding R-charges. In its simplest type IIB
nine-dimensional manifestation, which is also the most suited
for our purposes,
the partition function reads
\eqn\iibss{
\eqalign{
{\scr T} =& (V_8 \bar V_8 + S_8 \bar S_8 ) \varGamma^{(1,1)}_{m,2n} +
(O_8 \bar O_8 + C_8 \bar C_8) \varGamma^{(1,1)}_{m,2n+1}
\cr
&- (V_8 \bar S_8 + S_8 \bar V_8 ) \varGamma^{(1,1)}_{m+{1\over 2}, 2n}
- (O_8 \bar C_8 + C_8 \bar O_8 ) \varGamma^{(1,1)}_{m+{1\over 2}, 2n+1} \,,
\cr}
}
with $\varGamma^{(1,1)}_{m,n}$ the (1,1)-dimensional Narain lattice
associated to a circle of radius $R$, with Kaluza-Klein momentum $m$ and
winding $n$. For notational simplicity, the summation over momenta and
windings is suppressed, and we follow the notation of \AngelantonjCT.
$O_n$, $V_n$,
$S_n$ and $C_n$ are the ${\rm SO} (n)$ characters corresponding to the
identity, the vector, the spinor and the conjugate spinor representations.
In this string generalisation of the Scherk-Schwarz mechanism, all
space-time fermions with even windings
have evidently masses shifted by $1/2R$ as compared to
their would-be bosonic superpartners, while in the odd-winding sector
there is also a reversed GSO projection.

The model \iibss\ with supersymmetry breaking is still invariant under
world-sheet parity $\varOmega$ that, henceforth, can be gauged
\refs{\SagnottiTW}. We are
actually interested in T-dual IIA orientifold
models obtained by an $\varOmega
{\scr I}$ projection, where
${\scr I}$ is the inversion of the compact coordinate $x^9$. The
corresponding Klein-bottle amplitude reads
\eqn\kbss{
{\scr K} = {\textstyle{1\over 2}} (V_8 - S_8 ) W^{(1)}_{2n} +
{\textstyle{1\over 2}} (O_8 - C_8 ) W^{(1)}_{2n+1} \,,
}
and, consequently,
\eqn\mtbkt{
\Kt = {2^5 \over 2 R } \left( V_8 P^{(1)}_{2m} -
S_8 P^{(1)}_{2m+1} \right) \,,
}
where, as usual, the tilde in the amplitude stands for its
transverse-channel
({\it i.e.} closed-string) representation, and $P^{(1)}$, $W^{(1)}$ denote
the momentum or winding partition functions: $P^{(1)}_m \equiv
\varGamma^{(1,1)}_{m,0}$, $W^{(1)}_{n} \equiv \varGamma^{(1,1)}_{0,n}$.
{}From the expression \mtbkt\ one can immediately read the geometry of the
orientifold planes introduced by the $\varOmega{\scr I}$ projection
\DudasBN. Indeed,
after rewriting \mtbkt\ in the form
$$
\Kt = {2^5 \over 2 R} \left[ V_8 \left( {1 + (-1)^m \over 2}
\right)^2  - S_8 \left( {1 - (-1)^m \over 2} \right)^2 \right]
P^{(1)}_{m} \,,
$$
one can see that two orientifold 8-planes are sitting
at the edges of the segment
$S^1 /\bb{Z}_2$, {\it i.e.}
at the fixed points of the $\varOmega {\scr I}$ involution, and, moreover
the relative minus sign in the $S_8$ coefficient neatly reveals that
the two planes are CPT
conjugate, namely ${\scr O}^+$ and $\bar{\scr O}^+$.
Overall, one has the configuration of figure 1.

\vbox{
\centerline{\epsfbox{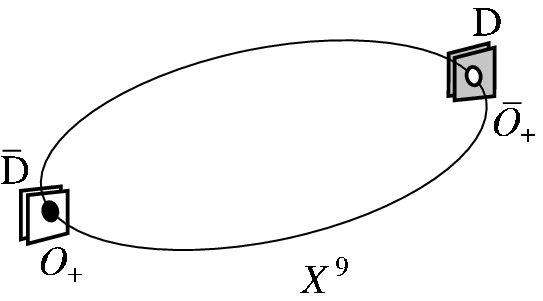}}
\centerline{Fig.1. The geometry of the nine-dimensional M-theory breaking
model}
}

As usual, tadpole cancellation demands the introduction of an open-string
sector that, in the case at hand, is bound to involve brane-antibrane
pairs. The annulus amplitude then reads
\eqn\ass{
{\scr A} = {\textstyle{1\over 2}} \left( N_{\rm D}^2 + N_{\bar{\rm D}}^2
\right) (V_8 - S_8) W^{(1)}_{n} + N_{\rm D} N_{\bar{\rm D}} (O_8 - C_8)
W_{n+{1\over 2}} \,,
}
where the first term describes open strings ending on a stack of
$N_{\rm D}$ branes or $N_{\bar{\rm D}}$ antibranes, while the second
term, with reverted GSO projection, pertains
to open strings stretched between branes and antibranes.
An $S$-modular transformation, brings the annulus amplitude in the
transverse-channel description
$$
\eqalign{
\At &= {2^{-5} \over 2 R} \left[ (N_{\rm D} + N_{\bar{\rm D}}
)^2 \left( V_8 P_{2m}^{(1)} - S_8 P^{(1)}_{2m+1} \right)
+ (N_{\rm D} - N_{\bar{\rm D}} )^2 \left( V_8 P_{2m+1}^{(1)} -
S_8 P^{(1)}_{2m} \right)\right]
\cr
&= {2^{-5}\over 2 R} \left[
\left( (-1)^m N_{\rm D} + N_{\bar{\rm D}} \right)^2 V_8 -
\left( (-1)^m N_{\rm D} - N_{\bar{\rm D}} \right)^2 S_8 \right]
P^{(1)}_{m} \,,
\cr}
$$
that, actually, makes transparent our choice of Wilson lines,
or brane displacements: we have here decided to put
the $N_{\rm D}$ branes on top of the $\bar{\scr O}^+$ plane at
$x^9 = \pi R$, while the $N_{\bar{\rm D}}$ antibranes are at $x^9 = 0$,
where an ${\scr O}^+$ plane lives \footnote{$^\dagger$}{Here we use
the conventions of \WittenBS:
an ${\scr O}^\pm$ plane has both negative (positive) tension and charge, while
a bar denotes its CPT conjugate with reversed R-R charge.}.

Finally, the transverse-channel M\"obius amplitude
$$
\Mt = - {1\over R} \left[ \left(
{1 + (-1)^m \over 2}\right) (N_{\rm D} + N_{\bar{\rm
D}} ) V_8 + \left(
{1 - (-1)^m \over 2}\right)(N_{\rm D} + N_{\bar{\rm D}} ) S_8 \right]
P_m^{(1)}
$$
is completely determined by $\Kt$ and $\At$, and in the
direct channel
\eqn\mss{
{\scr M} = - {\textstyle{1\over 2}} \left[ (N_{\rm D} + N_{\bar{\rm D}} ) V_8
+ (N_{\rm D} + N_{\bar{\rm D}}) S_8 (-1)^n \right] W^{(1)}_{n}
}
gives a proper (anti)symmetrisation of ${\scr A}$.

At the massless level one gets an ${\rm SO} (16) \otimes {\rm SO} (16)$
Chan-Paton gauge group with fermions in the $(136,1) \oplus (1,136)$
representation. In this open-string sector supersymmetry is broken
directly at the string scale as a consequence of our choice of Wilson
lines \AntoniadisGW. This breaking in the open-string sector
is closer in spirit to the ``brane supersymmetry
breaking'' mechanism of \refs{\SugimotoTX}
rather than to the Scherk-Schwarz reduction \AntoniadisKI.
The latter would correspond to exchange branes with antibranes, giving rise
to supersymmetric massless open-string excitations (brane supersymmetry).

The one-loop vacuum energy density results from the contributions of the
four surfaces of vanishing Euler character
$$
\eqalign{
\varLambda (R) =& {\textstyle{1\over 2}} \int_{\sscr F} {d^2 \tau \over
\tau_2^{11/2}} {{\scr T} (R) \over |\eta|^{14}} + \int_0^\infty
{d\tau_2 \over \tau_2^{11/2} } {{\scr K} (R) \over \eta^7 (2i\tau_2)}
\cr
&+ \int_0^\infty {d \tau_2 \over \tau_2^{11/2}} {{\scr A} (R) \over
\eta^7 ({i\tau_2 \over 2})} + \int_0^\infty {d\tau_2 \over
\tau_2^{11/2}} {{\scr M} (R) \over \eta^7 ( {1\over 2} + {i\tau_2 \over 2})}
\,,
\cr}
$$
where ${\scr F}$ is the familiar ${\rm SL} (2,\bb{Z})$
fundamental domain for the complex modulus $\tau$ of the world-sheet
torus. From the expressions \iibss , \kbss , \ass\ and \mss ,  it is then
clear that ${\scr T}$, ${\scr K}$
and ${\scr A}$ give rise to a cosmological constant which is power-law or
exponentially suppressed with the size of the compact Scherk-Schwarz
dimension, while this is not the case for
${\scr M}$ that yields, instead, a sizable
result set by the string scale.
This is a direct consequence of the miss-matching between
bosonic and fermionic degrees of freedom on the branes.
Actually, in the UV regime $\tau_2 \to 0$, ${\scr K}$, ${\scr A}$ and
${\scr M}$, independently, develop linear in $R$ contributions, related to
the emergence of local tadpoles.
Indeed, the general large-radius limit of the one-loop cosmological constant
in four dimensions is \refs{\AntoniadisGW,\marco}
\eqn\lambda{
\varLambda (R) \sim (n^{c}_{\rm B} - n^{c}_{\rm F} ) {1\over R^4} +
c_1 M_s^4 + c_2 (n^{o}_{\rm B} - n^{o}_{\rm F}) {M_s^{6-n} \over
R^{n-2}} + {\scr O} (e^{-M_s^2 R^2} ) \,,
}
where $n^c_{\rm B,F}$ and $n^o_{\rm B,F}$ denote the number of massless
bosonic and fermionic degrees of freedom from the closed and open string
sectors, respectively. The first term is a consequence of spontaneous
supersymmetry breaking by Scherk-Schwarz boundary conditions in the
closed-string sector
\refs{\ItoyamaRC, \Ldim, \AntoniadisZG},
while the second term is expected from the absence of supersymmetry on the
branes, leading to a non-vanishing localised energy density of order of the
string scale. Finally, the third term is a consequence of non-vanishing
local tadpoles of massless closed-string states, and its large radius
behaviour depends on the number $n$ of  dimensions transverse to the
branes that have a size comparable to $R$ \refs{\AntoniadisGW, \marco}.

Can we improve our model, associating to the closed-string spectrum of
\iibss\ and
\kbss\ an open-string sector still non-supersymmetric but with
Fermi-Bose degeneracy, at least at the massless level?
In other words, can we trade the $\bar{\scr O}^+$ plane at $x^9 = \pi R$
for an $\bar{\scr O}^-$ one, in order
to symmetrise the gauge bosons living on the
D-branes, while still anti-symmetrising the associated fermions?

\newsec{Introducing discrete deformations}

In a given model, the nature of the ${\scr O}$ planes content does not
depend solely on the type of the orientifold group ${\scr G}_\varOmega$, but
also on the values of the background fields. Indeed, a simple way to
transmute ${\scr O}$ planes consists in introducing discrete
deformations in the closed-string sector \WittenBS.
Although ${\scr G}_\varOmega$ projects out some of the
excitations of the parent theory, it was shown originally in \BianchiEU\ that
world-sheet parity is
still compatible with quantised values for their backgrounds. The
simplest instance where this phenomenon occurs is the compactification
on a two-dimensional torus with generic complex and K\"ahler structures.
The left-handed and right-handed momenta are then
$$
\eqalign{
p_{{\rm L},a} &= m_a + {1\over\alpha '} (g_{ab} - B_{ab} ) n^b \,,
\cr
p_{{\rm R},a} &= m_a - {1\over\alpha '} (g_{ab} + B_{ab} ) n^b \,.
\cr}
$$
Although asymmetric for generic values of the metric $g_{ab}$ and
the two-index antisymmetric tensor $B_{ab}$, the Narain
lattice can be made compatible with world-sheet parity if the NS-NS
two-form is quantised \BianchiEU
\eqn\quant{
B_{ab} = {\alpha ' \over 2} k \epsilon_{ab} \, ,
}
with $\epsilon_{ab}$ the Levi-Civita two-tensor and $k$ an integer.
Combined with the Peccei-Quinn translations, eq. \quant\ then identifies two
inequivalent classes:
$$
B_{ab} =0 \,, \qquad B_{ab} = {\alpha ' \over 2} \epsilon_{ab} \,.
$$
In orientifold
constructions, the first class corresponds to standard compactifications
(with vector structure) involving, in the T-dual version, four identical
${\scr O}^+$ planes sitting at the four fixed points of $T^2 /\bb{Z}_2$.
The second class, on the other hand, is
associated to $T^2$'s without vector structure, and has the
effect of halving the net charge and tension of the ${\scr O}$ planes
involved \WittenBS.
Indeed, the associated transverse-channel Klein-bottle amplitude
(in the case of an orthogonal torus with radii $R_1$ and $R_2$)
$$
\eqalign{
\Kt =& {2^3 \over 2} {\alpha ' \over R_1 R_2} (V_8 - S_8)
P_{(m_1 , m_2)}
\cr
&= {2^3 \over 2} {\alpha ' \over R_1 R_2} (V_8 - S_8)
\left( {1 - (-1)^{m_1} + (-1)^{m_2} + (-1)^{m_1 +m_2} \over 2} \right)^2
P_{(m_1 , m_2)}
\cr}
$$
clearly shows the origin of this reduction: among the four orientifold planes,
three have both negative tension and charge, {\it i.e.} they
are the familiar ${\scr O}^+$
planes, while the fourth one, at say $(x^1 , x^2) = (\pi R_1,0)$, has
reversed tension and charge, and is usually denoted as ${\scr O}^-$
plane (see fig. 2).

\vbox{
\vskip 0.5cm
\epsfxsize 10truecm
\centerline{\epsfbox{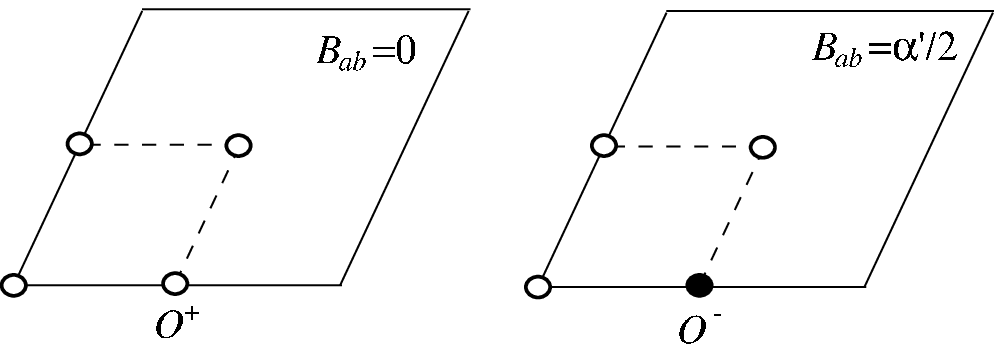}}
\centerline{Fig. 2. The ${\scr O}$ plane configuration for a non-trivial
$B_{ab}$ background}
\vskip 0.5cm
}

As a result, tadpole conditions now call for the introduction of only
sixteen D-branes (or eight ${\scr G}_\varOmega$ invariant combinations,
according to different ways of counting) and the associated gauge group
is orthogonal or symplectic depending on the nature of the orientifold
plane they are next to (${\scr O}^+$ or ${\scr O}^-$, respectively)
\BianchiEU.

\newsec{The model}

We have now got all the ingredients to concoct the model we are
looking for. One has simply to compactify  the nine-dimensional
model we described in section 2 on
an additional $T^2$ without vector structure,
{\it i.e.} with a non-vanishing $B$-field, and distribute the D-branes as
in figure 3.

\vbox{
\vskip 0.5cm
\epsfxsize 10truecm
\centerline{\epsfbox{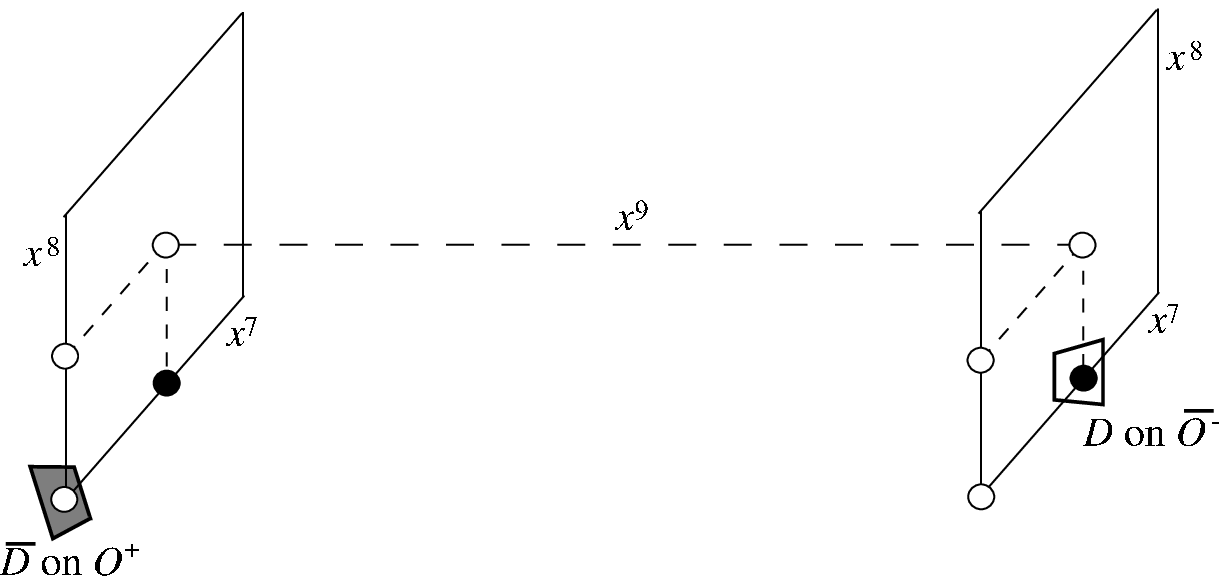}}
\centerline{Fig. 3. ${\scr O}$-planes and D-branes configurations
for the seven-dimensional orientifold.}
\vskip 0.5 cm
}

To be more concrete, we start from the type IIB model \iibss\ compactified on
a (2,2)-dimensional lattice, $\varGamma^{(2,2)} (B)$, with the NS-NS two-form
$B_{ab} = {\alpha ' \over 2} \epsilon_{ab}$,
$$
\eqalign{
{\scr T} =& \left[ (V_8 \bar V_8 + S_8 \bar S_8 ) \varGamma^{(1,1)}_{m,2n} +
(O_8 \bar O_8 + C_8 \bar C_8) \varGamma^{(1,1)}_{m,2n+1} \right.
\cr
& \left. - (V_8 \bar S_8 + S_8 \bar V_8 ) \varGamma^{(1,1)}_{m+{1\over 2}, 2n}
- (O_8 \bar C_8 + C_8 \bar O_8 ) \varGamma^{(1,1)}_{m+{1\over 2}, 2n+1}
\right] \varGamma^{(2,2)} (B) \,.
\cr}
$$
To fix the notation, $x^9$ is the coordinate on the circle of radius
$R$ responsible for the Scherk-Schwarz deformation, with $m$ and $n$ its
associated momenta and windings. Coordinates on the spectator $T^2$ are
labelled by the pair $(x^7 , x^8)$. For simplicity we choose $T^2$ to be
a squared torus with sides of size $2 \pi R_7$ and $2\pi R_8$, and,
of course, with a background $B_{ab} = {\alpha ' \over 2} \epsilon_{ab}$.
The corresponding momenta and windings are labelled by $(m_7 , m_8)$ and
$(n^7 , n^8)$.

Thus, we can immediately write down the Klein-bottle amplitude
$$
{\scr K} = {\textstyle{1\over 2}} \left[ (V_8 - S_8) W^{(1)}_{2n}
+ (O_8 - C_8) W^{(1)}_{2n+1} \right] W^{(2)}_{(2n^7 , 2 n^8)}
$$
associated to the $\varOmega {\scr I}$ orientifold projection, with now
${\scr I} x^i = - x^i$, an inversion on the three compact coordinates,
$i=7,8,9$.

The massless closed unoriented spectrum is simply given by the
dimensional reduction of that in section 2. In terms of ten-dimensional
fields it comprises the graviton, the dilaton and a R-R two-form potential.
For $R < \sqrt{\alpha '}$ a  tachyon appears in the spectrum
in the odd-winding (twisted) sector.
However, in this paper we are only interested in very large values for
$R$ and thus we are not concerned with this.

An $S$ modular transformation maps ${\scr K}$ into the transverse-channel
amplitude
$$
\eqalign{
\Kt =& {2^5 \over 2 \cdot 4} {(\alpha ' )^{3/2} \over
R R_8 R_7} \left[ V_8 \left( {1 + (-1)^m \over 2} \right)^2 - S_8
\left( {1-(-1)^m \over 2} \right)^2 \right] P_{m}^{(1)}
\cr
& \times \left( {1 - (-1)^{m_7} + (-1)^{m_8} + (-1)^{m_7 + m_8} \over 2}
\right)^2 P^{(2)}_{(m_7 , m_8)} \,.
\cr}
$$
As expected, two CPT-conjugate copies of the {\it r.h.s} two-torus of figure 2
are sitting at the edges of the $x^9$ segment, as shown in figure 3.

The rest of the construction is now straightforward. We have simply to
add eight brane-antibrane pairs to cancel the NS-NS tadpole in
$\Kt$. As usual, in the open sector we have the freedom to
distribute the branes along the compact dimensions.
In the case at hand, the configuration of figure 3 corresponds to
the annulus amplitude
$$
{\scr A} = {\textstyle{1\over 2}} \left( N_{\rm D}^2 + N_{\bar{\rm D}}^2
\right) (V_8 - S_8) W^{(1)}_{n} W^{(2)}_{(n^7 , n^8)}
+ N_{\rm D} N_{\bar{\rm D}} (O_8 - C_8 ) W^{(1)}_{n+{1\over 2}}
W^{(2)}_{(n^7 + {1\over 2}, n^8 )} \,,
$$
where, as before,
$N_{\rm D}$ ($N_{\bar{\rm D}}$) counts the number of branes (antibranes).

{}From
$$
\eqalign{
\At = & {2^{-5} \over 2} {(\alpha ' )^{3/2} \over R R_8 R_7}
\left[ \left( (-1)^m (-1)^{m_7} N_{\rm D} + N_{\bar{\rm D}} \right)^2 V_8
\right.
\cr
& \left. - \left( (-1)^m (-1)^{m_7} N_{\rm D} - N_{\bar{\rm D}} \right)^2 S_8
\right] P^{(1)}_{m} P^{(2)}_{(m_7 , m_8)}
\cr}
$$
and $\Kt$ we can immediately extract the M\"obius amplitude in the
transverse channel
$$
\eqalign{
\Mt = & - {1\over 2} {(\alpha ' )^{3/2} \over R R_8 R_7}
\left[ V_8 \left( {1+ (-1)^m \over 2} \right) + S_8 \left(
{1 - (-1)^m \over 2}\right) \right] P^{(1)}_{m}
\cr
& \times \left( {1 - (-1)^{m_7} + (-1)^{m_8} + (-1)^{m_7 + m_8} \over 2}
\right) \left( (-1)^{m_7} N_{\rm D} + N_{\bar{\rm D}} \right) P^{(2)}_{(m_7,
m_8 )} \,.
\cr}
$$
Upon a $P$-modular transformation, one then finds
$$
{\scr M} = - {\textstyle{1\over 2}} \left( V_8 + (-1)^n S_8 \right)
W^{(1)}_n \left[ (N_{\rm D} + N_{\bar{\rm D}} ) W^{(2)}_{(n^7 , 2 n^8 + 1)}
- (N_{\rm D} - N_{\bar{\rm D}} ) (-1)^{n^7} W^{(2)}_{(n^7 , 2 n^8)}
\right] \,,
$$
which is the proper (anti-)symmetrisation of ${\scr A}$.

The D-brane massless spectrum thus comprises the dimensional reduction
of ten-dimensional gauge bosons with gauge group ${\rm USp}(8)\otimes
{\rm SO}(8)$ and fermions in the $(28,1)\oplus (1,36)$ representations.
The two gauge group factors ({\it i.e.} the branes and the antibranes)
are located at different points along the compact $(x^8,x^9)$ directions,
as depicted in figure 3.
This spectrum is obviously Fermi-Bose degenerate, although
not supersymmetric. Moreover, in the limit $R,R_8\to\infty$
all massive string levels present Fermi-Bose degeneracy.
Actually, in this limit, the closed-string sector and the annulus become
supersymmetric, while supersymmetry is broken only by the M\"obius
amplitude, which reads
$$
{\scr M}\simeq {\textstyle{1\over 2}} (N_{\rm D} - N_{\bar{\rm D}})
(V_8 + S_8)(-1)^{n^7} W^{(1)}_{n^7}\, .
$$
This expression describes our choice of distributing branes
(antibranes) next to anti-orientifolds (orientifolds) planes,
thus breaking supersymmetry on their world-volumes.
Intuitively, this configuration is as if the fermionic superpartners
of the ${\rm USp}(8)$ and ${\rm SO}(8)$ gauge bosons, in the  $36$
and $28$ representations, were exchanged between the two sets of D-branes.

Note that supersymmetry is
restored in the limit $R_7 \ll \sqrt{\alpha'}$, which, in the
T-dual description, is equivalent to a large radius limit along
the brane world-volume, ${\tilde R}_7\equiv\alpha'/R_7 \gg \sqrt{\alpha'}$:
$$
{\scr M}\simeq {\textstyle{1\over 2}}
(N_{\rm D} - N_{\bar{\rm D}})\left[
(V_8 {\tilde P}^{(1)}_{2{\tilde m}_7}-
S_8{\tilde P}^{(1)}_{2{\tilde m}_7+1})-
(V_8 {\tilde P}^{(1)}_{2
{\tilde m}_7+1}-S_8{\tilde P}^{(1)}_{2{\tilde m}_7})\right]\, ,
$$
where ${\tilde P}^{(1)}$ denotes the momentum lattice in terms of the
dual radius ${\tilde R}_7$. This implies that, in this limit, the second
term in \lambda\ is to be replaced by $c_1(n^{o}_{\rm B} - n^{o}_{\rm F})
/{\tilde R}_7^4$, in analogy with the first term of the closed-string
sector, as a consequence of the Scherk-Schwarz supersymmetry breaking.
However, since gauge interactions are not localised along this direction,
${\tilde R}_7$ cannot become very large, otherwise the gauge theory
on the branes would be non perturbative.

\newsec{The vacuum energy density}

The one-loop vacuum energy receives contributions from the four
surfaces of vanishing Euler character
$$
\eqalign{
\varLambda (R) =& {\textstyle{1\over 2}} \int_{ {\sscr F}} {d^2 \tau \over
\tau_2^{9/2}} \, {{\scr T} (R) \over |\eta |^{10}} +
\int_0^\infty {d\tau_2 \over \tau_2^{9/2}} \, {{\scr K} (R) \over \eta^5}
\cr
&+ \int_0^\infty {d\tau_2 \over \tau_2^{9/2}} \, {{\scr A} (R) \over \eta^5}
+ \int_0^\infty {d\tau_2 \over \tau_2^{9/2}} \, {{\scr M} (R) \over
\hat\eta ^5} \,,
\cr}
$$
where ${\scr T} (R)$, ${\scr K} (R)$, ${\scr A} (R)$ and ${\scr M} (R)$ are
given in the previous section. Using the Jacobi identity
$$
V_8 - S_8 = {\textstyle{1\over 2}} \left( {\vartheta_3^4
\over \eta^4 } - {\vartheta^4_4 \over \eta^4} - {\vartheta^4_2 \over \eta^4}
\right) \equiv 0\,,
$$
and the tadpole condition $N_{\rm D} = N_{\bar{\rm D}} = 8$, one can cast
the various contributions in the simpler form
$$
{\scr T} (R) = \left[ \left| {\vartheta_2 \over \eta}\right|^8 \, (-1)^m
\varGamma^{(1,1)}_{m,n} + \left| {\vartheta_4 \over \eta }\right|^8 \,
\varGamma^{(1,1)}_{m,n+{1\over 2}} + \left| {\vartheta_3 \over \eta }\right|^8
\, (-1)^m \varGamma^{(1,1)}_{m,n+{1\over 2}} \right] \varGamma^{(2,2)}
(B) \,,
$$
$$
{\scr K} (R) = {\textstyle{1\over 2}}
{\vartheta_4^4 (0|2i\tau_2) \over \eta^{7} (2i\tau_2)}
\, \sum_{n,n^7,n^8} e^{-{\pi\tau_2 \over \alpha '}
\left[ (2n+1)^2 R^2 + (2n^7 R_7)^2 + (2n^8 R_8)^2 \right]}
 \,,
$$
$$
{\scr A} (R) = 64 {\vartheta_4^4 \bigl( 0 \big| {i\tau_2 \over 2} \bigr)
\over \eta^{7} \bigl( {i\tau_2 \over 2} \bigr)}
\, \sum_{n,n^7,n^8} e^{-{\pi\tau_2 \over \alpha '}
\left[ (n+{1\over 2} )^2 R^2 + (n^7 +{1\over 2})^2 R_7^2 +
(n^8 R_8 )^2 \right]} \,,
$$
and
$$
{\scr M} (R) = -8 {\hat\vartheta_2^4 \bigl( 0 \big| {1\over 2}+
{i\tau_2 \over 2} \bigr)
\over \hat\eta^{7} \bigl( {1\over 2}+
{i\tau_2 \over 2} \bigr)} \,
\, \sum_{n,n^7,n^8} e^{-{\pi\tau_2 \over \alpha '}
\left[ (2nR)^2 + (n^7 R_7)^2 + (2n^8 +1)^2 R_8^2 \right]} \,.
$$
It is then clear that, for $\tau_2 > 0$, the contributions from the
Klein-bottle, annulus and M\"obius-strip amplitudes can be made arbitrarily
small by taking two of the compact directions, $R$ and $R_8$, very large,
since
$$
\varLambda ({\scr K}) +
\varLambda ({\scr A}) +
\varLambda ({\scr M}) \sim {\scr O} (e^{-R^2} \ {\rm or}\  e^{-R_8^2}) \,.
$$

On the other hand, the potentially divergent region $\tau_2 =0$ is
better described, as usual, in terms of the horizontal time $\ell \sim
1/\tau_2$. The amplitudes then read
$$
\eqalign{
\Kt =& {2 (\alpha ')^{3/2} \over RR_7 R_8}
\int_0^\infty d\ell \, {\vartheta_2^4 (0|i\ell) \over
\eta^{12} (i\ell)}\, \sum (-1)^m \, e^{-{\pi \ell \alpha ' \over 2}
\left[ \left( {m\over R} \right)^2 + \left( {m_7 \over R_7}\right)^2
+\left( {m_8 \over R_8}\right)^2 \right]} \,,
\cr
\At =& {2 (\alpha ')^{3/2} \over RR_7 R_8}
\int_0^\infty d\ell \, {\vartheta_2^4 (0|i\ell) \over
\eta^{12} (i\ell)}\, \sum (-1)^{m+m_7} \, e^{-{\pi \ell \alpha ' \over 2}
\left[ \left( {m\over R} \right)^2 + \left( {m_7 \over R_7}\right)^2
+\left( {m_8 \over R_8}\right)^2 \right]} \,,
\cr
\Mt =& -{4 (\alpha ')^{3/2} \over RR_7 R_8}
\int_0^\infty d\ell {\hat\vartheta_2^4 (0|i\ell+
{1\over 2}) \over
\eta^{12} (i\ell +{1\over 2})}\,
\sum (-1)^{m_8} \,
e^{-{\pi \ell \alpha ' \over 2}
\left[ \left( {m\over R} \right)^2 + \left( {2m_7 \over R_7}\right)^2
+\left( {m_8 \over R_8}\right)^2 \right]} \,,
\cr}
$$
where we have explicitly shown the integration over the transverse-channel
modulus $\ell$.
In the large radius limit, the integrals are dominated by the
closed-string channel infra-red region
$\ell \to \infty$. It is therefore appropriate to rescale the horizontal
time as $\alpha ' \ell / R^2$. As a result, the massive
string excitations decouple
in the large radius limit and the amplitudes reduce to
\eqn\uvres{
\eqalign{
\Kt =& {32 R \sqrt{\alpha'}  \over R_7 R_8} \int_0^\infty d\ell\,
\sum (-1)^m \, e^{-{\pi \ell \over 2}
\left[ m^2 + \left({R \over R_7}\right)^2
m_7^2 + \left({R \over R_8}\right)^2 m_8^2 \right]} \,,
\cr
\At =& {32 R \sqrt{\alpha'} \over R_7 R_8} \int_0^\infty d\ell\,
\sum (-1)^{m + m_7} \,
e^{-{\pi \ell \over 2} \left[ m^2 + \left({R \over R_7}\right)^2
m_7^2 + \left({R \over R_8}\right)^2 m_8^2 \right]} \,,
\cr
\Mt =& - {64 R \sqrt{\alpha'} \over R_7 R_8} \int_0^\infty d\ell\,
\sum (-1)^{m_8} \,
e^{-{\pi \ell \over 2} \left[ m^2 + 4\left({R \over R_7}\right)^2
m_7^2 + \left({R \over R_8}\right)^2 m_8^2 \right]} \,.
\cr}
}
It is then evident that for $R\sim R_8 \gg R_7$ only $m_7 =0 $ survives,
and the contribution of
$\Kt$, $\At$ and $\Mt$ to the vacuum energy becomes
\eqn\uvsum{
\eqalign{
\Kt +\At +\Mt =& {64R\sqrt{\alpha'}\over R_7 R_8}\int_0^\infty d\ell\,
\sum \left( (-1)^m - (-1)^{m_8}\right) \,
e^{-{\pi\ell\over 2}\left[ m^2 +\left({R\over R_8}\right)^2 m_8^2\right]}
\cr
\simeq& {128\over\pi}{R\sqrt{\alpha'}\over R_7 R_8}\sum
{(-1)^m - (-1)^{m_8}\over m^2 +\left({R\over R_8}\right)^2 m_8^2}\, ,
\cr}
}
and vanishes in the symmetric case $R=R_8$, up to exponentially small
corrections, as stated before.

\vbox{
\vskip 0.5cm
\epsfxsize 7truecm
\centerline{\epsfbox{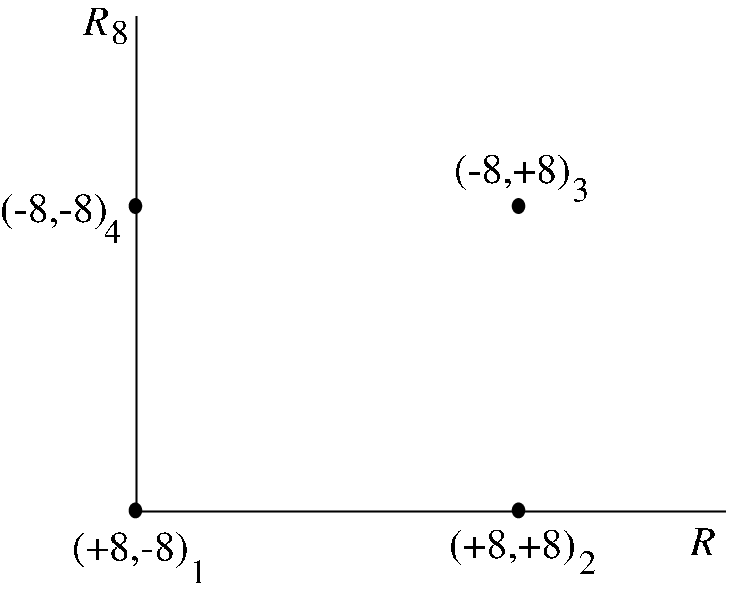}}
\centerline{Fig. 4. Mass and charge distribution in the $R, R_8 \gg R_7$
limit.}
\vskip 0.5 cm
}

Actually, these amplitudes can be given a suggestive geometrical
interpretation
in terms of gravitational and ``Maxwell''-type interactions
\PolchinskiMT, where
the condition $R=R_8$ would translate into an overall compensation of forces
(a BPS-like condition). In the limit $R_7 \ll R, R_8$,
we are mainly left with a
two-dimensional distribution of charged and massive point-like sources
as in figure
4. Here, the two entries $(M_i \,,\, q_i)$ in the parenthesis refer to the
mass and charge of the $i$-th ``particle'', where in our conventions a single
${\scr O}7^{+}$ plane has both tension and charge equal to $-4$ while a D
brane carries a (positive) unit of tension and charge. As a result,
the overall
gravitational force experienced by the system is
$$
\eqalign{
F_{\rm Newton}
=& - (M_1 M_2 + M_3 M_4 ) \, \log \, R - (M_1 M_4 + M_2 M_3 ) \, \log \,
R_8
\cr
& - (M_1 M_3 + M_2 M_4 ) \, \log \, \sqrt{R^2 + R^2_8}
\cr
=& - 128 \, \log \, R + 128 \, \log \, R_8 + 128 \,
\log \, \sqrt{R^2 + R^2_8} \,.
\cr}
$$
Similarly, the ``Maxwell'' force due to the R-R field exchange reads
$$
\eqalign{
F_{\rm Maxwell}
&= + (q_1 q_2 + q_3 q_4 ) \, \log \, R + (q_1 q_4 + q_2 q_3 ) \, \log \,
R_8 + (q_1 q_3 + q_2 q_4 ) \, \log \, \sqrt{R^2 + R^2_8}
\cr
&= - 128 \, \log \, R + 128 \, \log \, R_8 - 128 \,
\log \, \sqrt{R^2 + R^2_8} \,,
\cr}
$$
and thus the total force is
$$
F_{\rm Newton} + F_{\rm Maxwell} = - 256 \, \log \, R + 256 \, \log \, R_8 \,.
$$
It is then clear that, as stated previously, the condition $R= R_8$, required
to cancel the leading contribution to the cosmological constant, translates
into a (BPS-like) zero-force condition on the distribution of D-branes and
${\scr O}$-planes.

The analysis of the torus amplitude is standard.
The twisted (odd-winding)
sector does not have massless Kaluza-Klein states and thus gives
an exponentially suppressed contribution. On the contrary, the
untwisted sector contributes with the familiar power-law fall-off $1/R^7$.
Upon compactification to four dimensions ({\it e.g.}
on spectator tori) one recovers
the expected $1/R^4$ behaviour.

\newsec{A six-dimensional example with small cosmological constant}

We can now proceed to describe lower-dimensional models with a naturally
small one-loop cosmological constant. As before, they consist of suitable
combinations of Scherk-Schwarz reductions \AntoniadisKI\ and non-trivial
discrete deformations \AngelantonjJH.

In general, lower-dimensional orientifolds have a rich structure due to
the simultaneous presence of D9 and D5 branes (or T-duals of them), on which
Scherk-Schwarz deformations can act differently. However, the proper
combination of discrete deformations and modified boundary conditions
should guarantee the existence of a non-supersymmetric spectrum with
(sets of) branes with Fermi-Bose
degeneracy at all massive levels in the large radius limit.
As we saw before, this is needed for suppressing the vacuum energy
in the gauge (brane) sector.

In six dimensions one could start with the $T^4 /\bb{Z}_2$ orbifold model
\AntoniadisKI
$$
\eqalign{
{\scr T} =& {\textstyle{1\over 2}} \Bigl\{ \left( |V_4 O_4 + O_4 V_4 |^2
+ |S_4 S_4 + C_4 C_4 |^2 \right) \varGamma_{m,2n}
\cr
&+ \left( |O_4 O_4 + V_4 V_4 |^2 + |S_4 C_4 + C_4 S_4 |^2 \right)
\varGamma_{m,
2n +1}
\cr
&- \left[ (V_4 O_4 + O_4 V_4 ) (\bar S_ 4 \bar S_ 4 + \bar C _4 \bar C_ 4
) + {\rm h.c.} \right] \varGamma_{m+{1\over 2}, 2n}
\cr
& - \left[ (O_4 O_4 + V_4 V_4 ) (\bar S_4 \bar C _4 + \bar C _4 \bar S_4)
+ {\rm h.c.} \right] \varGamma_{m+{1\over 2},2n+1} \Bigr\} \varGamma^{(1,1)}
\varGamma^{(2,2)} (B)
\cr
& + {\textstyle{1\over 2}} \left( |V_4 O_4 - O_4 V_4 |^2 + |S_4 S_4 - C_4 C_4
|^2 \right) \left| {2\eta \over \vartheta_2}\right|^4
\cr
& + {\textstyle{16\over 4}} \left(
|O_4 C_4 + V_4 S_4 - S_4 O_4 - C_4 V_4 |^2 +
|O_4 S_4 + V_4 C_4 - C_4 O_4 - S_4 V_4 |^2 \right) \left| {\eta \over
\vartheta_4}\right|^4
\cr
&+ {\textstyle{16\over 4}} \left(
|O_4 C_4 - V_4 S_4 - S_4 O_4 + C_4 V_4 |^2 +
|O_4 S_4 - V_4 C_4 - C_4 O_4 + S_4 V_4 |^2 \right) \left| {\eta \over
\vartheta_3}\right|^4 \,,
\cr}
$$
where now the internal lattice has been deformed by the inclusion of a
non-trivial $B$-field along a $T^2$. As expected, ``untwisted'' massless
fermions
(and, in particular, the gravitini) have acquired a mass proportional to
the inverse compactification radius. On the contrary, the ``twisted''
ones stay massless, since they are localised at fixed points and thus
do not feel the Scherk-Schwarz deformation. Notice also that the twisted
sector is now non chiral as a result of modular invariance.

Following \AngelantonjJH, the presence of a
discrete $B$-field acts as a projector onto even-winding contributions in
${\scr K}$, while no restriction is imposed on the momenta excitations,
so that
$$
\eqalign{
{\scr K} =& {\textstyle{1\over 4}} \left( V_4 O_4 + O_4 V_4 -
S_4 S_4 - C_4 C_4
\right) \left( P^{(4)}_{(m_1 , m_2 , m_3 , m_4)} + W^{(4)}_{(2 n_1 , 2n_2 ,
n_3 , 2n_4)} \right)
\cr
& + {\textstyle{1\over 4}} \left( O_4 O_4 + V_4 V_4 - S_4 C_4 - C_4 S_4
\right) W^{(4)}_{(2n_1 , 2n_2 , n_3, 2n_4 +1)}
\cr
&+ 2 \Bigl[ (O_4 C_4 + V_4 S_4 -S_4 O_4 - C_4 V_4 ) + (O_4 S_4 + V_4 C_4 -
C_4 O_4 - S_4 V_4 ) \Bigr] \left( {\eta \over \vartheta_4}\right)^2
\cr}
$$
is a proper (anti-)symmetrisation of the torus amplitude.
Here, $(m_1 , m_2)$ and $(n_1 , n_2)$ denote momenta and windings on
the $\varGamma^{(2,2)} (B)$ lattice, while $m_4$ and $n_4$ refer to the
``Scherk-Schwarz'' direction. The projected massless spectrum is also non
chiral and comprises the graviton, one (unconstrained) two-form and seventeen
scalars from the untwisted sector, together with
52 scalars, two (unconstrained) two-forms and eight Dirac fermions
from the twisted sector.

The transverse-channel amplitude
reveals that while the ${\scr O}9$ plane is a conventional one, the
${\scr O}5$ planes reproduce the geometry described in section 4 (with an
additional spectator $S^1$). Thus, the open string sector involves both
D9 branes and D5 brane-antibrane pairs. The corresponding one-loop amplitudes
can again be retrieved from \AntoniadisKI\ and \AngelantonjJH, and read
$$
\eqalign{
{\scr A}_{99} =& {\textstyle{1\over 4}} \bb{I}_{\rm N}^2
(V_4 O_4 + O_4 V_4) \sum_{a,b=0,{1\over 2}}
P^{(4)}_{(m_1 +a , m_2+b , m_3 , m_4)}
\cr
& - {\textstyle{1\over 4}}
\bb{I}^2_{\rm N} (S_4 S_4 + C_4 C_4 ) \sum_{a,b=0,{1\over 2}}
P^{(4)}_{(m_1 +a, m_2+b  , m_3 , m_4+{1\over 2})}
\cr
&+ {\textstyle{1\over 4}}
R_{\rm N}^2 (V_4 O_4 - O_4 V_4 ) \left( {2\eta \over \vartheta_2}
\right)^2 \,,
\cr}
$$
in the D9-D9 sector,
$$
\eqalign{
{\scr A}_{55} =& {\textstyle{1\over 4}} \bb{I}_{\rm D}^2 (V_4 O_4 + O_4 V_4
-S_4 S_4 - C_4 C_4 ) W^{(4)}_{(n_1 , n_2 , n_3 ,n_4)}
\cr
& + {\textstyle{1\over 4}} R^2_{\rm D} (V_4 O_4 - O_4 V_4
+ S_4 S_4 - C_4 C_4 ) \left( {2\eta\over \vartheta_2}\right)^2 \,,
\cr}
$$
in the D5-D5 sector,
$$
\eqalign{
{\scr A}_{\bar 5 \bar 5} =&
{\textstyle{1\over 4}} \bb{I}_{\bar{\rm D}}^2 (V_4 O_4 + O_4 V_4
-S_4 S_4 - C_4 C_4 ) W^{(4)}_{(n_1 , n_2 , n_3 ,n_4)}
\cr
& + {\textstyle{1\over 4}} R^2_{\bar{\rm D}} (V_4 O_4 - O_4 V_4
- S_4 S_4 + C_4 C_4 ) \left( {2\eta\over \vartheta_2}\right)^2 \,,
\cr}
$$
in the ${\rm D}\bar 5$-${\rm D}\bar 5$ sector, and
$$
\eqalign{
{\scr A}_{5\bar 5} =&
{\textstyle{1\over 2}} \bb{I}_{\rm D} \bb{I}_{\bar{\rm D}}
(O_4 O_4 + V_4 V_4 - S_4 C_4 - C_4 S_4 ) W^{(4)}_{(n_1 +{1\over 2}
, n_2 , n_3 , n_4 +
{1\over 2})} \,,
\cr
{\scr A}_{95} =& \bb{I}_{\rm N} \bb{I}_{\rm D} (O_4 C_4 + V_4 S_4 - S_4 O_4
- C_4 V_4 ) \left( {\eta \over \vartheta_4} \right)^2
\cr
&+i R_{\rm N} R_{\rm D} (O_4 C_4 - V_4 S_4 -S_4 O_4 + C_4 V_4 )
\left( {\eta \over \vartheta_3} \right)^2 \,,
\cr
{\scr A}_{9\bar 5} =&
\bb{I}_{\rm N} \bb{I}_{\bar{\rm D}} (O_4 S_4 + V_4 C_4 - C_4 O_4
- S_4 V_4 ) \left( {\eta \over \vartheta_4} \right)^2
\cr
&+ R_{\rm N} R_{\bar{\rm D}} (O_4 S_4 - V_4 C_4 -C_4 O_4 + S_4 V_4 )
\left( {\eta \over \vartheta_3} \right)^2 \,,
\cr}
$$
in the mixed ones.

Finally, the M\"obius-strip amplitudes
$$
\eqalign{
{\scr M}_9 =& - {\textstyle{1\over 4}} \bb{I}_{\rm N} \Biggl[
(V_4 O_4 + O_4 V_4 ) \sum_{a,b=0,{1\over 2}} (-1)^{2a (2b-1)}
 P^{(4)}_{(m_1 + a,
m_2+b , m_3 , m_4)}
\cr
&-(S_4 S_4 + C_4 C_4) \sum_{a,b=0,{1\over 2}} (-1)^{2a (2b-1)}
P^{(4)}_{(m_1 + a,
m_2+b , m_3 , m_4+{1\over 2})}
\cr
&- (V_4 O_4 - O_4 V_4 ) \left( {2\eta \over\vartheta_2}\right)^2 \Biggr] \,,
\cr}
$$
$$
\eqalign{
{\scr M}_5 &= -{\textstyle{1\over 4}} \bb{I}_{\rm D} \Biggl\{
\left[ V_4 O_4 + O_4 V_4 +(-1)^{n_4} \left( S_4 S_4 + C_4 C_4 \right) \right]
\cr
&\times
\left( W_{(n_1, 2n_2 +1 , n_3 n_4)}^{(4)} - (-1)^{n_1}
W^{(4)}_{(n_1 , 2 n_2 , n_3 , n_4)} \right)
\cr
& -(V_4 O_4 - O_4 V_4 - S_4 S_4 + C_4 C_4 ) \left( {2\eta \over \vartheta_2}
\right)^2 \Biggr\} \,,
\cr}
$$
and
$$
\eqalign{
{\scr M}_{\bar 5} &= -{\textstyle{1\over 4}} \bb{I}_{\bar {\rm D}} \Biggl\{
\left[ V_4 O_4 + O_4 V_4 +(-1)^{n_4} \left( S_4 S_4 + C_4 C_4 \right) \right]
\cr
&\times \left( W_{(n_1, 2n_2 +1 , n_3 n_4)}^{(4)} + (-1)^{n_1}
W^{(4)}_{(n_1 , 2 n_2 , n_3 , n_4)} \right)
\cr
& -(V_4 O_4 - O_4 V_4 + S_4 S_4 - C_4 C_4 ) \left( {2\eta \over \vartheta_2}
\right)^2 \Biggr\} \,,
\cr}
$$
which as usual are obtained by the factorisation of $\Kt$ and $\At$,
complete the orientifold projection in the open-string sector, and
suggest the following parametrisation of Chan-Paton charges
$$
\eqalign{
\bb{I}_{\rm N} &= n + \bar n \,,
\cr
\bb{I}_{\rm D} &= d_1 +d_2 \,,
\cr
\bb{I}_{\bar{\rm D}} &= d + \bar d \,,
\cr}
\qquad\quad
\eqalign{
R_{\rm N} &= i (n-\bar n) \,,
\cr
R_{\rm D} &= d_1 - d_2 \,,
\cr
R_{\bar{\rm D}} &= i (d-\bar d) \,.
\cr}
$$

As usual, tadpole conditions fix the overall number of branes
$$
\bb{I}_{\rm N} = 16 \,, \quad \bb{I}_{\rm D} =8 \,, \quad
\bb{I}_{\bar{\rm D}} = 8 \,,
$$
that is again reduced by the presence of the discrete deformation
\AngelantonjJH.

The massless (anomaly-free) spectrum, obtained by the expansion of
${\scr A}$ and ${\scr M}$
$$
\eqalign{
{\scr A}^{(0)} \sim & \left[ n\bar n + {\textstyle{1\over 2}} (d_1^2 + d_2^2)
+ d \bar d \right] V_4 O_4
+ \left[ {\textstyle{1\over 2}} (n^2 + \bar n^2) + d_1 d_2 +
{\textstyle{1\over 2}} (d^2 + \bar d^2) \right] O_4 V_4
\cr
&- {\textstyle{1\over 2}} \left( d_1^2 + d_2^2 + d^2 + \bar d^2 \right)
C_4 C_4
- (d_1 d_2 + d\bar d) S_4 S_4
\cr
& + 2 (n d_2 + \bar n d_1 ) (O_4 C_4 - S_4 O_4 )
+ 2 (n\bar d + \bar n d) (O_4 S_4 - C_4 O_4 ) \,,
\cr
{\scr M}^{(0)} \sim & {\textstyle{1\over 2}} \left[ (d_1 + d_2) V_4 O_4
- (n+ \bar n + d + \bar d) O_4 V_4 + (d_1 + d_2 - d - \bar d ) C_4 C_4 \right]
\,,
\cr
}
$$
comprises a $\left[ {\rm U}(8) \right]_9 \otimes \left[ {\rm USp} (4) \otimes
{\rm USp} (4) \right]_5 \otimes \left[ {\rm U} (4) \right]_{\bar 5}$
Chan-Paton gauge group with
$$
\eqalign{
\hbox{\rm four scalars} :& \
(28 \oplus \overline{28} ; 1,1;1 ) \oplus (1;4,4;1) \oplus (1;1,1;6
\oplus \bar 6)
\cr
\hbox{\rm four scalars} :& \ (8;1,4;1) \oplus (\bar 8;4,1;1)
\oplus (8;1,1;\bar 4) \oplus
(\bar 8;1,1; 4)
\cr
\hbox{\rm one left-handed spinor} :& \ (1;6,1;1) \oplus (1;1,6;1) \oplus
(1;1,1; 10 \oplus \overline{10} )
\cr
& \ \oplus (8;1,1;\bar 4) \oplus (\bar 8;1,1;4)
\cr
\hbox{\rm one right-handed spinor} :& \ (1;4,4;1) \oplus (1;1,1;16)
\oplus (8;1,4;1) \oplus (\bar 8;4,1;1)
\cr}
$$
This spectrum neatly reveals that the mechanism we have described in the
previous sections is realised on the D5 system, while the D9 branes feel
the familiar  Scherk-Schwarz breaking \AntoniadisKI.
As a result, in the large radius
limit, the  one-loop cosmological constant decays like $1/R^4$ in the
bulk and on the D9 branes (upon further compactification to four
dimensions), while it is exponentially suppressed on the D5 branes due to
a Fermi-Bose degenerate spectrum.

\newsec{Conclusions and discussions}

We have here presented an explicit example
of non-supersymmetric orientifolds
with two large transverse dimensions and a naturally small
cosmological constant. In the large radius limit,
$R=R_8\to\infty$, supersymmetry is restored in the bulk, while the
D-brane spectra stay non-supersymmetric, though exhibiting Fermi-Bose
degeneracy at all massive string levels. This is the first instance of
a vacuum configuration
with non-supersymmetric non-abelian gauge sector and a naturally
small cosmological constant. In the
simplest case, the model contains two ``mirror worlds", while the
degeneracy is due to an exchange of ordinary superpartners on
the two branes.

On the other hand, brane supersymmetry is recovered when another
radius (${\tilde R}_7$), now longitudinal to the brane
world-volume, becomes larger than the string scale. However, ${\tilde
R}_7$ cannot be made arbitrarily large, since gauge
interactions propagate along this direction, and thus, they would become
strongly coupled. In the absence of Fermi-Bose
degeneracy, this would imply a power-law fall-off of the brane
cosmological constant, whose magnitude would be determined, in this
case, by the size of this large dimension instead of the string scale.

The main question that remains to be addressed concerns, of course,
higher-order corrections. In the following, we
give general arguments in favour of their weakness based on the main
features of the model presented in this paper.
In the large radius limit,  $R=R_8\to\infty$, and for fixed
world-sheet modulus, the only sizable contribution to $\varLambda$ comes
from non-BPS interactions between coincident boundaries (D-branes)
and crosscaps (${\scr O}$-planes)\footnote{$^\dagger$}{In fact,
the non-supersymmetric ({\it i.e.} non-vanishing) brane/brane
and ${\scr O}$-plane/${\scr O}$-plane interactions are exponentially
suppressed in $R$ and $R_8$.}.
At the one-loop level, such interactions are described by one diagram, the
M\"obius strip, and the vacuum energy receives two
contributions from $\bar{\rm D}$-branes/${\scr O}^+$-plane and
D-branes/$\bar{\scr O}^-$-plane interactions. Since the projection of the
two orientifold planes, which locally breaks supersymmetry, is opposite in
the two ``mirror worlds'', the two contributions to the vacuum energy,
equal in magnitude, have opposite signs, thus yielding a vanishing result.

A similar mechanism is therefore expected to take place in higher-genus
surfaces containing zero or one crosscap, while a problem might
emerge for surfaces containing two crosscaps,
together with any boundary.\footnote{$^\ddagger$}{Indeed, a general
classification of non-oriented and topologically inequivalent Riemann
surfaces implies that there are only three classes of diagrams
characterised by 0, 1 or 2 crosscaps and an arbitrary number of holes and
handles.} The simplest surface of this type arises at genus 3/2,
and has one hole and two crosscaps. Again, one has to sum over
the two contributions associated to the
two different sets of ``massless'' brane/${\scr O}$-plane configurations.
A preliminary analysis of such a diagram indicates indeed that the two
contributions are equal but now come with even sign, yielding a brane
vacuum energy of the order of $M_s^4$. However, in the limit ${\tilde
R}_7\gg \ell_s$, this contribution is expected to
vanish, since supersymmetry is recovered.
Hence, one is left with potential contributions from
various degeneration limits of the world-sheet
modular parameters (period matrix).

To better elucidate this point, let us summarise here the
one-loop case studied in detail in sections 4 and 5. In the
ultraviolet limit, $\tau_2\to 0$, that is not in the fundamental domain
for the modulus of the torus, the (transverse-channel) Klein-bottle,
annulus and M\"obius amplitudes contribute as
$$
\eqalign{
\Kt\! +\!\At\! +\!\Mt\!\simeq& {64{\tilde R}_7\over R R_8}\sum_{m,m_8}
\left[ \left( 1+{N_{\rm D}N_{\bar{\rm D}}\over 64}\right) (-1)^m -
{N_{\rm D}+N_{\bar{\rm D}}\over 8} (-1)^{m_8}\right]
\int_0^\infty d\ell
e^{-{\pi\ell\over 2}\left( {m^2\over R^2} +
{m_8^2\over R_8^2}\right)}
\cr
=& {64{\tilde R}_7\over R^2}
\left( 1-{N_{\rm D}\over 8}\right)
\left( 1-{N_{\bar{\rm D}}\over 8}\right)
\sum_{m,m_8} (-1)^m\int_0^\infty d\ell
e^{-{\pi\ell\over 2}{m^2 + m_8^2\over R^2}}\, ,
\cr}
$$
where, for convenience, we have restored the Chan-Paton factors for
branes ($N_{\rm D}$) and antibranes
($N_{\bar{\rm D}}$), located next to the $\bar{\scr O}^-$ and
${\scr O}^+$ planes, respectively.
Clearly, the ultraviolet contribution is due to the sum
of local tadpoles of all massless string states, and vanishes if such
tadpoles are absent, {\it i.e.} when $N_{\rm D}=N_{\bar{\rm D}}=8$.

On the other hand, the infrared region, $\tau_2\to\infty$, may
receive contributions only from zero-winding sectors.
The large radius
behaviour is now determined by the restoration of supersymmetry.
In the closed-string
sector, one finds the familiar contribution $(n_B^c-n_F^c)/R^4$ which is
suppressed below the experimental limit for extra dimensions of
sub-millimetre size.
In the open-string sector, however, one gets a term proportional to
$(n_B^o-n_F^o)/{\tilde R}_7^4$, which can not be sufficiently dumped
since ${\tilde R}_7^{-1}$ is in the TeV region. More precisely, following
our general analysis in section 4, the M\"obius amplitude contributes with
$$
{\scr M}\simeq 4 (N_{\rm D} - N_{\bar{\rm D}})
\int_0^\infty{dt\over t^3}\sum_{n^7}
e^{-{\pi\over 2t}(2n^7+1)^2{\tilde R}_7^2}\, \sim\,
{N_{\rm D} - N_{\bar{\rm D}}\over {\tilde R}_7^4}\, ,
$$
However, because of Fermi-Bose degeneracy in the D-brane spectrum
$(N_{\rm D}=N_{\bar{\rm D}}=8)$, this term is absent and the next
correction is exponentially suppressed in
${\tilde R}_7\gg\ell_s$.

Note the similarity of these
models with the heterotic duals
of Fermi-Bose degenerate closed-string vacua of ref. \HarveyRC,
where the
inverse of the heterotic radius is proportional to the type IIA string
coupling constant, and supersymmetry is restored in the large radius limit.
This similarity, together with the property of local tadpole cancellation
in the vacuum energy, suggests that higher order corrections could be
suppressed, at least in a class of models satisfying additional
constraints, along the lines of the closed-string examples \sp.
In any case, the precise analysis of higher-genus
corrections and possible extra conditions in our constructions remains an
interesting open problem.

\bigskip
\noindent
{\bf Acknowledgement} We thank Tom Taylor for enlightening discussions.
C.A. would like to acknowledge the Physics Department of Oxford University
for hospitality during the completion of this work.
This work was supported in part by the European Commission under the RTN
contract HPRN-CT-2000-00148, the INTAS contract 55-1-590 and the Royal
Society.

\listrefs

\end